\documentclass[letterpaper]{article}

\usepackage{natbib,alifeconf}  
\usepackage{booktabs}
\usepackage{amsmath, amssymb, amsfonts}
\usepackage{url,hyperref,cleveref}
\usepackage{standalone}
\usepackage{tikz}
\usetikzlibrary{calc,arrows.meta,fit,positioning}
\tikzset{
  inputnode/.style={
    circle,draw,minimum size=12pt,inner sep=1pt
  },
  hexnode/.style={
    circle,draw,minimum size=12pt,inner sep=0pt,fill=white
  },
  box/.style={
    draw,rounded corners,inner sep=6pt
  },
  hexarrow/.style={
    ->,>=stealth,
    line width=0.4pt,
    shorten >=8pt,shorten <=5pt
  },
  dashedarrow/.style={
    ->,dashed,>=stealth,
    line width=0.5pt,
    shorten >=5pt,   
    shorten <=4pt    
  }
}

\title{Cognition without neurons: modelling anticipation in a basal reservoir computer}

\author{
    Polyphony Bruna$^{1*}$  \and
    Linnéa Gyllingberg$^{2*}$ \\
    \mbox{}\\
    $^1$Department of Cognitive and Information Sciences, University of California, Merced, USA\\
    $^2$Department of Mathematics, University of California, Los Angeles, USA\\
    $^*$Corresponding authors: pbruna@ucmerced.edu, linnea@math.ucla.edu
} 

\begin{document}

\maketitle

\begin{abstract}
    How do non-neural organisms, such as the slime mould \textit{Physarum polycephalum}, anticipate periodic events in their environment? We present a minimal, biologically inspired reservoir model that demonstrates simple temporal anticipation without neurons, spikes, or trained readouts. The model consists of a spatially embedded hexagonal network in which nodes regulate their energy through local, allostatic adaptation. Input perturbations shape energy dynamics over time, allowing the system to internalize temporal regularities into its structure. After being exposed to a periodic input signal, the model spontaneously re-enacts those dynamics even in the absence of further input -- a form of unsupervised temporal pattern completion. This behaviour emerges from internal homeodynamic regulation, without supervised learning or symbolic processing. Our results show that simple homeodynamic regulation can support unsupervised prediction, suggesting a pathway to memory and anticipation in basal organisms. 
\end{abstract}

Submission type: \textbf{Full Paper}\\

Data/Code available at: https://github.com/basal-reservoir


\section{Introduction}

How do simple organisms, without brains, centralised nervous systems, or even neurons, store information, anticipate change, and respond meaningfully to patterns in their environment? These questions are central to a growing field of research into minimal (sometimes called "basal") cognition \citep{lyon2021reframing, van2006principles, levin2019computational} that aims to identify the fundamental conditions under which cognitive phenomena can emerge. 

A canonical model organism in this field is the acellular slime mould, \textit{Physarum polycephalum}, known for a wide range of cognitive abilities, such as habituation \citep{boisseau2016habituation}, anticipation of periodic stimuli \citep{saigusa2008amoebae}, and solving shortest path problems \citep{nakagaki2000maze, beekman2015brainless}; i.e., finding the shortest path between two food sources. These behaviours emerge in the absence of a nervous system, suggesting that complex responses can arise through decentralised, embodied mechanisms. In \textit{Physarum polycephalum}, memory has been proposed to be stored directly in the organism’s body through tubes connecting nutrient sources that thicken with repeated use, encoding past experience in morphology \citep{kramar2021encoding, boussard2021adaptive}. As a result, the body adapts to reflect structural features of its environment, enabling it to adjust its behaviour in advance of future states as a form of anticipation \citep{dubois2003anticipation, steppturvey2009}. These behaviours challenge the assumption that anticipation requires neural systems and instead point to anticipatory processes that are found across biological systems -- including microbes, plants, and even gene regulatory networks -- that all operate without neurons or brains \citep{deans2021biological, pla2025minimal}.

These observations have prompted broader theoretical reflections on how cognition might arise in non-neural systems. In minimal or basal cognition, cognitive capacities are seen as emerging from the dynamics of self-organizing, adaptive systems rather than explicit computation \citep{lyon2021reframing, van2006principles}. Recent philosophical work further refines this view by arguing that cognitive memory can only arise within a metabolic individual \citep{sims2024slime} — that is, a self-sustaining, homeostatically regulated system that actively maintains its own organization \citep{ godfrey2013darwinian}. On this account, morphological or physiological changes only qualify as cognitive memory if they participate in an ongoing regulatory loop that maintains the system’s organization.  

This perspective is echoed in the ecological psychology tradition \citep{gibson1955perceptual, gibson1972direct, michaels2014tencommandments, turvey2012onintelligence} in cognitive science, which endeavours to explain various cognitive and adaptive behaviours using perception-action loops and other local regulatory mechanisms to account for how organisms directly "resonate" with the informational structure of their environment without the need for a representational intermediary. These mechanisms suggest a path towards intelligent behaviour based not on internalized prediction but on continuously adapting local action to the informational structure of one's environment. 

Ecological psychology departs from traditional cognitivist approaches that explain behaviour through explicit computations performed on internalized representations an organism carries of their environment. Insights from ecological psychology have been useful for explaining sensorimotor control systems, such as how insects manage in-flight navigation around obstacles \citep{srinivasan1992opticflow} and how humans catch flying projectiles \citep{fink2009flyball}, demonstrating that local regulatory mechanisms can explain coordination dynamics not only within basal cognition but in complex, neural cognition as well.

Several modelling approaches have explored how non-neural systems can display adaptive or learning-like behaviour, though often with differing goals. For example, \cite{beer2004autopoiesis} has explored in depth the autopoietic self-sustaining capabilities of emergent structures in the cellular automata of Conway’s Game of Life. Similarly, \cite{kondepudi2020dissipative} report symmetry-breaking and self-repairing capabilities of metallic beads exposed to electrical fields. Moreover, \cite{snezhko2011magnetic} have demonstrated coordination and self-assembly among nickel microspheres exposed to magnetic field pulses. The slime mould \textit{Physarum polycephalum} has played a particularly prominent role in demonstrations of basal cognitive capabilities \citep{adamatzky2010physarummachines}.

Indeed, computer simulations of \textit{Physarum polycephalum} have been especially informative. Agent-based models have simulated foraging and network formation \citep{jones2010characteristics}, while flow-based and reaction–diffusion models have described how contractile dynamics shape efficient transport structures \citep{tero2010rules}. More recently, oscillatory and flow-based dynamics have been combined to model learning and decision-making processes within the slime mould \citep{gyllingberg2025minimal}. These models focus on explaining \textit{Physarum polychephalum}'s behaviour directly, often emphasizing morphological adaptation to environmental conditions. By contrast, electronic analogies have also been proposed: For example, \citet{pershin2009memristive} modelled \textit{Physarum} anticipation using a memristor-based LC circuit that replicated learning-like dynamics, though without spatial structure or self-organizing adaptation.

In parallel, research on physical learning has investigated how soft materials or mechanical systems can solve learning tasks through local learning rules \citep{stern2020supervised, stern2023learning}. However, most physical learning models remain grounded in supervised learning frameworks, meaning that the system adapts by minimizing an explicit reward function that compares the system’s output to a desired target, often requiring external supervision or optimization.  In contrast, in non-neural biological cognition there is no obvious reward function or task-specific output to optimize. Instead, behaviour emerges from ongoing regulation and coordination within the organism itself. For example, the skin-brain thesis proposes that early nervous systems evolved not to optimize input–output mappings but to coordinate movement across excitable tissues \citep{keijzer2013nervous}. This perspective shifts the focus from optimization to the self-organized dynamics of the body as the basis for adaptive behaviour.

Recent theoretical work has proposed that rich internal dynamics may evolve under selection for flexible behaviour. Seoane \citeyearpar{seoane2019evolutionary} argues that reservoir computing architectures — where a fixed recurrent network separates and propagates signals through time while a simple readout extracts outputs — could emerge naturally in biological systems. However, classical reservoir computing still assumes a separation between internal dynamics and output decoding \citep{jaeger2001echo, maass2002real}. In organisms like \textit{Physarum polycephalum}, cognitive behaviour appears to arise directly from internal regulation without a distinct readout.

Closer to this idea, Falandays et al. \citeyearpar{falandays2021prediction, falandays2024resonance} reframed prediction not as inference in an input-output system but as multiscale pattern completion through time using local, metabolically inspired mechanisms in a spiking recurrent network motivated by reservoir computing. Their model exhibits predictive behaviour through internal, allostatic adaptation of the reservoir weights without the explicit representation and minimization of prediction error. The authors argue that predictive processing in neural systems like humans can emerge from pattern completion through time in self-organizing, distributed systems without the need for an explicit predictive mechanism.

In this paper, we extend a Falandays-style network by removing the spiking mechanism and introducing a spatially structured reservoir. We propose a state-continuous, non-neural adaptive reservoir inspired by the decentralised organization of \textit{Physarum}, which we refer to as a basal reservoir computer. The network is arranged as a hexagonal lattice inspired by the organism’s tubular morphology and typical degree of connectivity \citep{dussutour2024flow}. Much like the Falandays network, instead of a fixed reservoir with trained readouts typical of a reservoir computer, the basal reservoir network adapts locally through allostatic dynamics that adjust node weights with neighbours over time to maintain desired levels of energy in each node. The result is a model that implicitly learns to anticipate perturbations in the temporal structure of its environment through internal, local regulation without explicit prediction error minimization.

Our goal is not to mimic the computational capacity of neural learning in a basal reservoir but to explore how adaptive regulation can give rise to memory and prediction in minimal biological organisms. This view echoes work on basal cognition \citep{bich2016role}, where regulatory feedback enables organisms to maintain internal coherence while responding flexibly to environmental signals. By drawing on general computational principles -- how systems encode past experience into structure and re-enact it through dynamics -- we show how anticipatory behaviour can emerge in a non-neural system through local, unsupervised adaptation. In our model, this occurs via distributed interactions between a spatially structured body and its environment, echoing the behaviour of slime moulds.

\section{Model}

\begin{figure}[ht]
  \centering
  \begin{tikzpicture}[x=1cm,y=1cm]

  \begin{scope}[shift={(4,0)},local bounding box=hexgrid]
    \def\R{1} \def\NC{2} \def\NR{2} \def\delta{0.07}
    \foreach \i in {0,...,\NC}{
      \foreach \j in {0,...,\NR}{
        \pgfmathsetmacro\xc{1.5*\i*\R}
        \pgfmathsetmacro\yc{sqrt(3)*\j*\R + mod(\i,2)*sqrt(3)/2*\R}
        \coordinate (A\i\j) at ($(\xc,\yc)+(  0:\R)$);
        \coordinate (B\i\j) at ($(\xc,\yc)+( 60:\R)$);
        \coordinate (C\i\j) at ($(\xc,\yc)+(120:\R)$);
        \coordinate (D\i\j) at ($(\xc,\yc)+(180:\R)$);
        \coordinate (E\i\j) at ($(\xc,\yc)+(240:\R)$);
        \coordinate (F\i\j) at ($(\xc,\yc)+(300:\R)$);
        \foreach \u/\v/\ang in {
          A\i\j/B\i\j/60, B\i\j/C\i\j/120,
          C\i\j/D\i\j/180, D\i\j/E\i\j/240,
          E\i\j/F\i\j/300, F\i\j/A\i\j/0
        }{
          \pgfmathsetmacro\dxA{\delta*cos(\ang+90)}
          \pgfmathsetmacro\dyA{\delta*sin(\ang+90)}
          \pgfmathsetmacro\dxB{\delta*cos(\ang-90)}
          \pgfmathsetmacro\dyB{\delta*sin(\ang-90)}
          \draw[hexarrow,transform canvas={shift={(\dxA,\dyA)}}]
            (\u)--(\v);
          \draw[hexarrow,transform canvas={shift={(\dxB,\dyB)}}]
            (\v)--(\u);
        }
        \foreach \pt in {A\i\j,B\i\j,C\i\j,D\i\j,E\i\j,F\i\j}
          \node[hexnode] at (\pt) {};
      }
    }
  \end{scope}

  \node[box,fit=(hexgrid)] (reservoirBox) {};
  \node[above=2pt of reservoirBox.north] {Reservoir layer};

  \coordinate (inCenter)
    at ($(reservoirBox.center)+(-5cm,0)$);
  \node[inputnode] (i1) at ([yshift=+2cm] inCenter) {$i_1$};
  \node[inputnode] (i2) at ([yshift=+1cm] inCenter) {$i_2$};
  \node[inputnode] (i3) at ([yshift= 0cm] inCenter)  {$i_3$};
  \node          (dots) at ([yshift=-1cm] inCenter)  {$\vdots$};
  \node[inputnode] (iM) at ([yshift=-2cm] inCenter) {$i_M$};
  \node[box,fit=(i1)(i2)(i3)(dots)(iM)] (inputBox) {};
  \node[above=2pt of inputBox.north] {Input layer};

  \draw[dashedarrow] (i1) to[bend left=20]  (B02);
  \draw[dashedarrow] (i1) to[bend right=15] (C01);

  \draw[dashedarrow] (i2) to[bend left=15]  (C11);
  \draw[dashedarrow] (i2) to[bend right=10] (D10);

  \draw[dashedarrow] (i3) to[bend left=10]  (D20);
  \draw[dashedarrow] (iM) to[bend right=18] (E21);

\end{tikzpicture}
  \caption{%
    Schematic of the basal reservoir computer architecture. \textbf{Input layer} (left): $M$ input nodes $i_1,\dots,i_M$ deliver external signals into the reservoir via dashed, directed arrows representing input weights $W^{\mathrm{in}}_{n,m}$. \textbf{Reservoir layer} (right): $N$ processing nodes arranged on a hexagonal lattice. Between each neighbouring pair of nodes $n$ and $n'$ there are two solid, parallel arrows — one from $n$ to $n'$ (weight $W_{n',n}$) and one from $n'$ to $n$ (weight $W_{n,n'}$) — to indicate bidirectional energy flow (in general, $W_{n,n'}\neq W_{n',n}$). No readout layer is included: all computation and memory emerge from the reservoir’s internal dynamics.}
  \label{fig:basal-reservoir-schematic}
\end{figure}
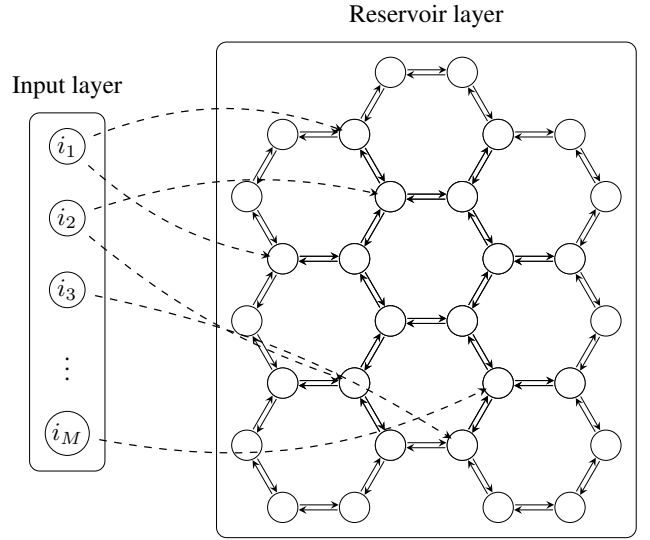

Our model is inspired by experimental observations of \textit{Physarum polycephalum} where the organism appears to anticipate periodic environmental changes \citep{saigusa2008amoebae}. However, we do not attempt to model any specific experimental system or biological mechanism in detail. Instead, we take inspiration from the broader phenomenon -- adaptive, anticipatory behaviour without neurons -- to explore how a non-spiking network can achieve similar capabilities through local, unsupervised, and allostatic mechanisms. The basic architecture is shown in Figure \ref{fig:basal-reservoir-schematic}.

\subsubsection{Model Structure}
The model consists of two components:
\begin{enumerate}
    \item an \textbf{input layer} composed of $M$ input nodes, and
    \item a \textbf{reservoir layer} composed of $N$ processing nodes arranged in a hexagonal lattice.
\end{enumerate}

The input layer is represented as a set of nodes $\{i_1, \ldots, i_M\}$ that deliver external input signals to the reservoir. Connections from the input layer to the reservoir are defined by an input weight matrix $W^{\mathrm{in}} \in \mathbb{R}^{N \times M}$, where $W^{\mathrm{in}}_{n,m}$ represents the strength of the connection from input node $i_m$ to reservoir node $n$. Input nodes are externally driven, and their activations $I_m(t)$ perturb the reservoir dynamics at each time step $t$.

The reservoir layer is formally represented as a directed, weighted graph $G = (V, E)$, where $V = \{1, \ldots, N\}$ is the set of nodes and $E \subseteq V \times V$ is the set of directed edges. Each edge $(n', n) \in E$ carries a weight $W_{n,n'}(t)$ representing the strength of energy flow from node $n'$ to node $n$ at time $t$. All initial weights are drawn from a Gaussian distribution with mean $\mu=0$ and standard deviation $\sigma=1$. The reservoir nodes are spatially embedded as a hexagonal lattice, and each node has, on average, three neighbouring nodes, inspired by the network structure observed in slime mould networks \citep{dussutour2024flow}.

Each node in the reservoir layer, $n \in V$, is characterized by a current energy level $X_n(t)$ (initialized at 0) and a dynamic target energy level $T_n(t)$ (initialized at 1). The adaptation of these values over time is governed in the following manner.

\subsubsection{Reservoir Activation Dynamics}
At each time step $t$ the activation vector of the reservoir is updated based on: 
\begin{enumerate}
    \item external input from input nodes (weighted by $W^{\mathrm{in}}_{n,m}$),
    \item endogenous flow from neighbouring reservoir nodes (weighted by $W_{n,n'}$),
    \item and passive leak of energy, $\lambda$.
\end{enumerate}

The activation update rule is given by:

\begin{equation}
X_n(t) = \sum_{m=1}^{M} W^{\mathrm{in}}_{n,m} I_m(t) + \sum_{n' \in \mathcal{N}(n)} \lambda  W_{n,n'}(t) X_{n'}(t-1),
\end{equation}

where $\mathcal{N}(n)$ denotes the set of neighbours of node $n$ in the reservoir.

\subsubsection{Homeodynamic Adaptation and Learning Rules}

At each time step, nodes adjust both the weight of their incoming connections ($W_{n,n'}(t)$) -- i.e., nodes from which energy flowed -- and their own target energy level ($T_n(t)$) based on their energetic state. The core principle is allostasis: each node seeks to maintain an energy level close to its internal target. The energetic error of each node is calculated as follows:

\begin{equation}
E_n(t) = X_n(t) - T_n(t).
\end{equation}

Adaptation follows two simple rules:
\begin{enumerate}
\item If $E_n(t) < 0$ (energy too low), the node \textbf{strengthens} its incoming connections and \textbf{lowers} its target energy.
    \item If $E_n(t) > 0$ (energy too high), the node \textbf{weakens} its incoming connections and \textbf{raises} its target energy.
\end{enumerate}

(If $E_n(t) = 0$, weights and targets \textbf{do not change}.)\\

This learning rule can be interpreted as a simple form of metabolic regulation: nodes adjust both their "demand" for energy and their "connectivity" to their environment in order to maintain stability under fluctuating external inputs. In biological terms, this process is loosely analogous to how slime moulds regulate the flow of resources through their tubules. Here it is framed more generally as an adaptive, homeodynamic process. These updates provide the reservoir with a rudimentary form of metabolic self-regulation, making the network as a whole a ‘metabolic individual’ in the sense of \cite{godfrey2013darwinian} and \cite{sims2024slime}. 

Incoming connection weights are updated based on the energetic error and the relative contribution of active neighbouring nodes. Specifically, we define $\hat{X}_{n'}(t)$ as the normalized activation of neighbour $n'$ relative to the total incoming activation to node $n$:

\begin{equation}
\hat{X}_{n'}(t) = \frac{X_{n'}(t) W_{n,n'}(t)}{\sum_{k \in \mathcal{N}(n)} X_k(t) W_{n,k}(t)}.
\end{equation}
This term captures the proportion of total incoming energy to node $n$ attributable to neighbour $n'$. The weight update rule is as follows: 

\begin{equation}
W_{n,n'}(t+1) = W_{n,n'}(t) - \eta_W \, E_n(t) \, \hat{X}_{n'}(t),
\end{equation}
where $\eta_W$ is the weight learning rate.

Lastly, each node also adapts its target energy level:
\begin{equation}
T_n(t+1) = T_n(t) + \eta_T E_n(t),
\end{equation}
where $\eta_T$ is the target learning rate. 

These local learning rules allow the reservoir to embed information about the temporal structure of external inputs into its internal organization. Rather than relying on supervised learning or explicit readouts, the system reconfigures its internal energy flow through local, unsupervised, and homeodynamic regulation. In the following section, we demonstrate that this minimal model can exhibit anticipation through self-organized and distributed internal dynamics.

\section{Results}

Inspired by simple anticipatory slime mould studies \cite{saigusa2008amoebae, mayne2017coupled}, in which light or temperature systematically oscillates between two values, $M=1$ input node was fully connected to $N=48$ reservoir nodes. The model was perturbed with a simple periodic signal that oscillated between input values of 0 ("OFF") and 1 ("ON") over 500 time steps. Every node in the reservoir received the same input value. $\lambda$, $\eta_W$, and $\eta_T$ were all fixed at 0.01.

\begin{figure}[h!]

    \centering
    \includegraphics[width=1\linewidth]{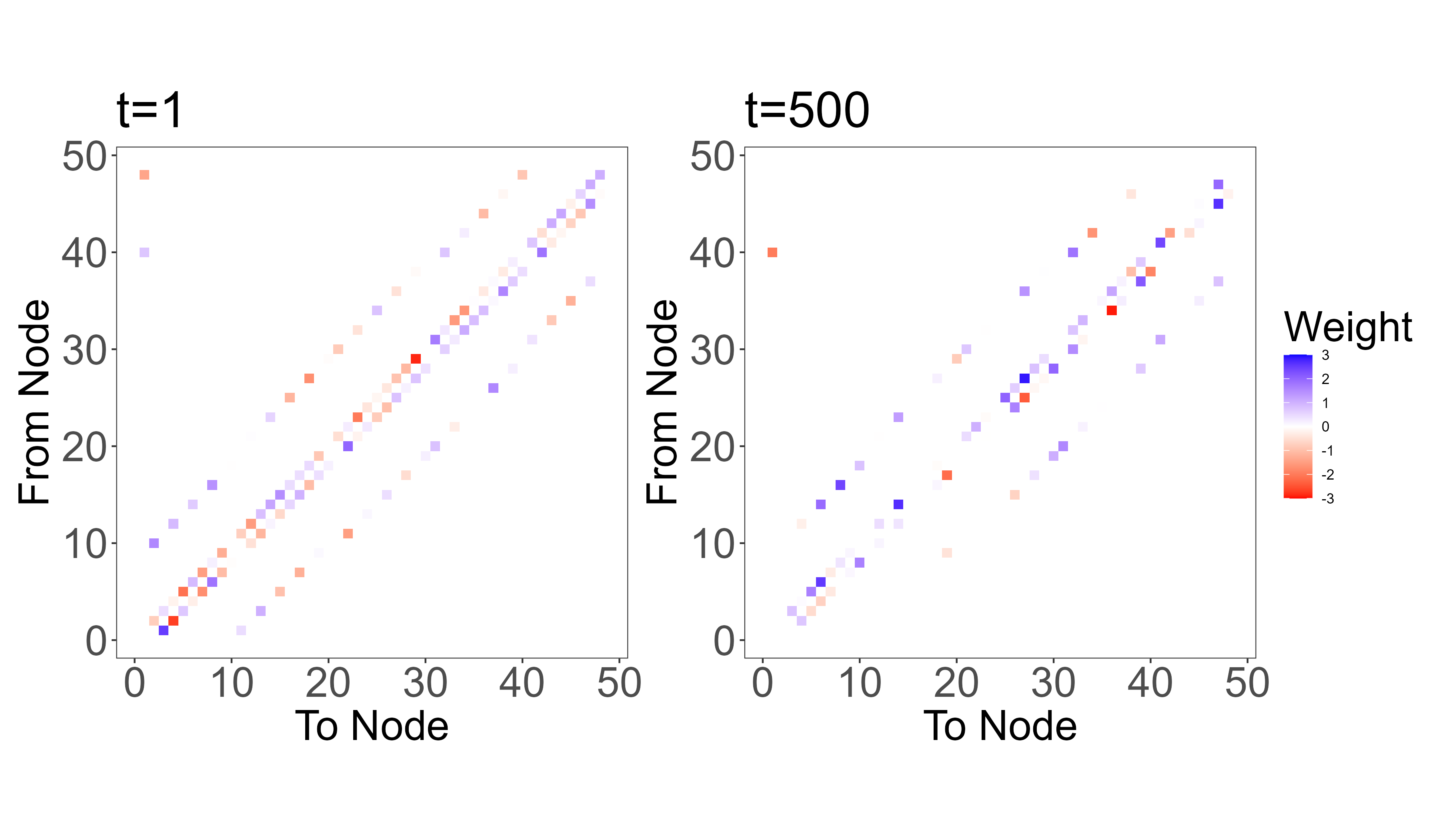}
    \caption{Reservoir weights on the first and last time step of training for a sample simulation. Weights are randomly sampled from a Gaussian distribution at initialization. Over the course of stimulus exposure, the weights adapt following local, homeodynamic regulation of activation, resulting in pruning and increased polarization of weight values.}
    \label{fig:weightsprepost}
\end{figure}

\begin{figure}[h!]
    \centering
    \includegraphics[width=1\linewidth]{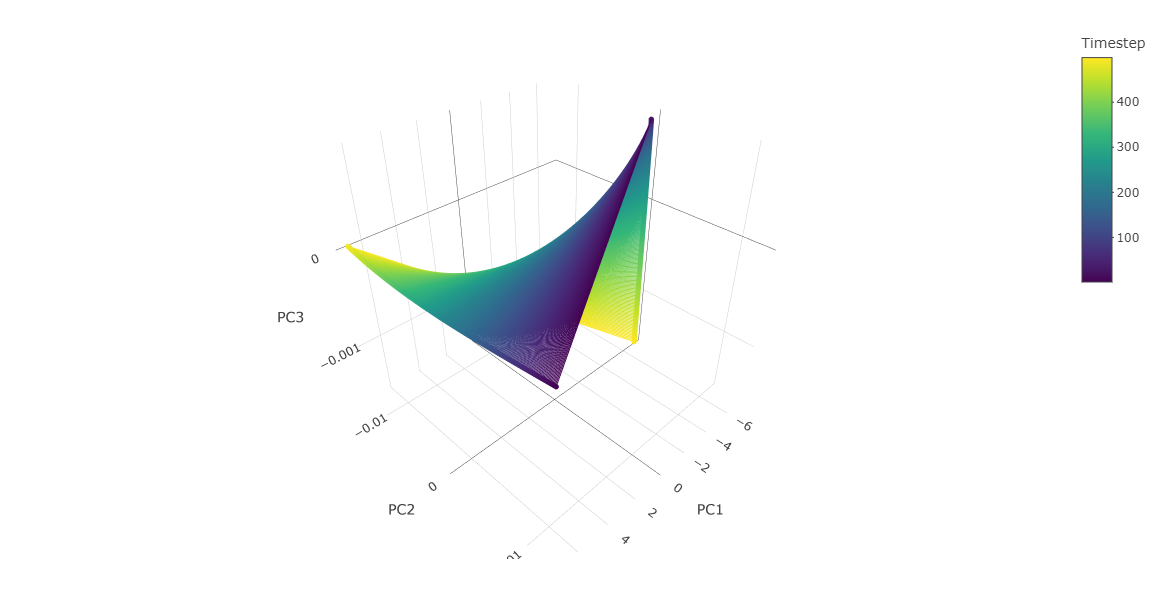}
    \includegraphics[width=0.75\linewidth]{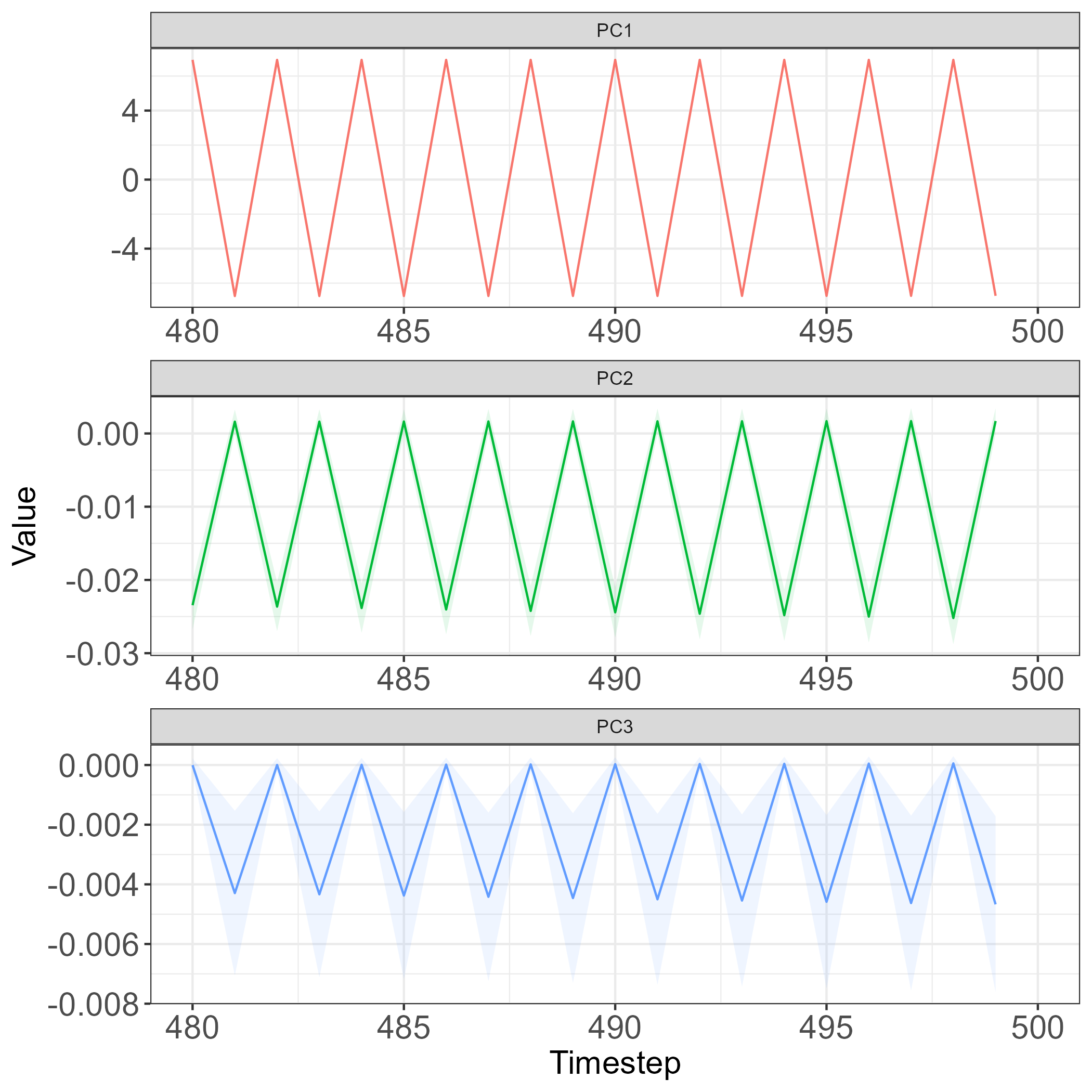}
    \caption{Reservoir state space during stimulus exposure. \textbf{Activation manifold} (top): Principal components analysis (PCA) of the reservoir's activation state space over training for a sample simulation. Reservoir activation exhibited a coherent routine, oscillating across a partition along the first principal component (PC1). Reservoir activity drifted through the state space over time while maintaining separation between two attractors corresponding to each possible state. \textbf{Response over time} (bottom): The first three principal components plotted over time for the last 20 time steps of training, averaged over 100 simulations. Very little variance in the oscillatory behaviour of PC1 indicates that across simulations the network successfully exhibited a stable response to the oscillatory signal. Greater variance along the second and third principal components is indicative of the drift of the oscillatory trajectory shown in the top figure, illustrating that the model maintains a stable behavioural response (PC1) to the signal despite underlying variance of the activation states producing the response over time.}
    \label{fig:pca}
\end{figure}

The reservoir weights before and after training for a sample simulation are shown in Figure \ref{fig:weightsprepost}. Weights were pruned over time and remaining connections polarized as the reservoir responded to the oscillatory signal. A principal component analysis (PCA) of the activation states corresponding to this simulation show that the reservoir dynamics exhibited a coherent but unstable routine during stimulus exposure (Figure \ref{fig:pca}), indicated by a periodic oscillation along the first principal component between two attractors in the reservoir state space. Plotting the first three principal components further illustrates that these two attractors drifted as exposure to the oscillatory signal increased while maintaining maximal separation between the two points.

After 500 time steps, the reservoir dynamics were carried forward without further perturbation (i.e., the reservoir only received inputs with a value of zero) to produce a fading memory signature. If the reservoir successfully learned to anticipate the periodic structure of the training regime, then the dynamics during the fading memory should reflect the structure of the training signal. In other words, the fading memory of the reservoir should carry the oscillatory signal forward in time.

\begin{figure*}[h!]
    \centering
    \includegraphics[width=1\linewidth]{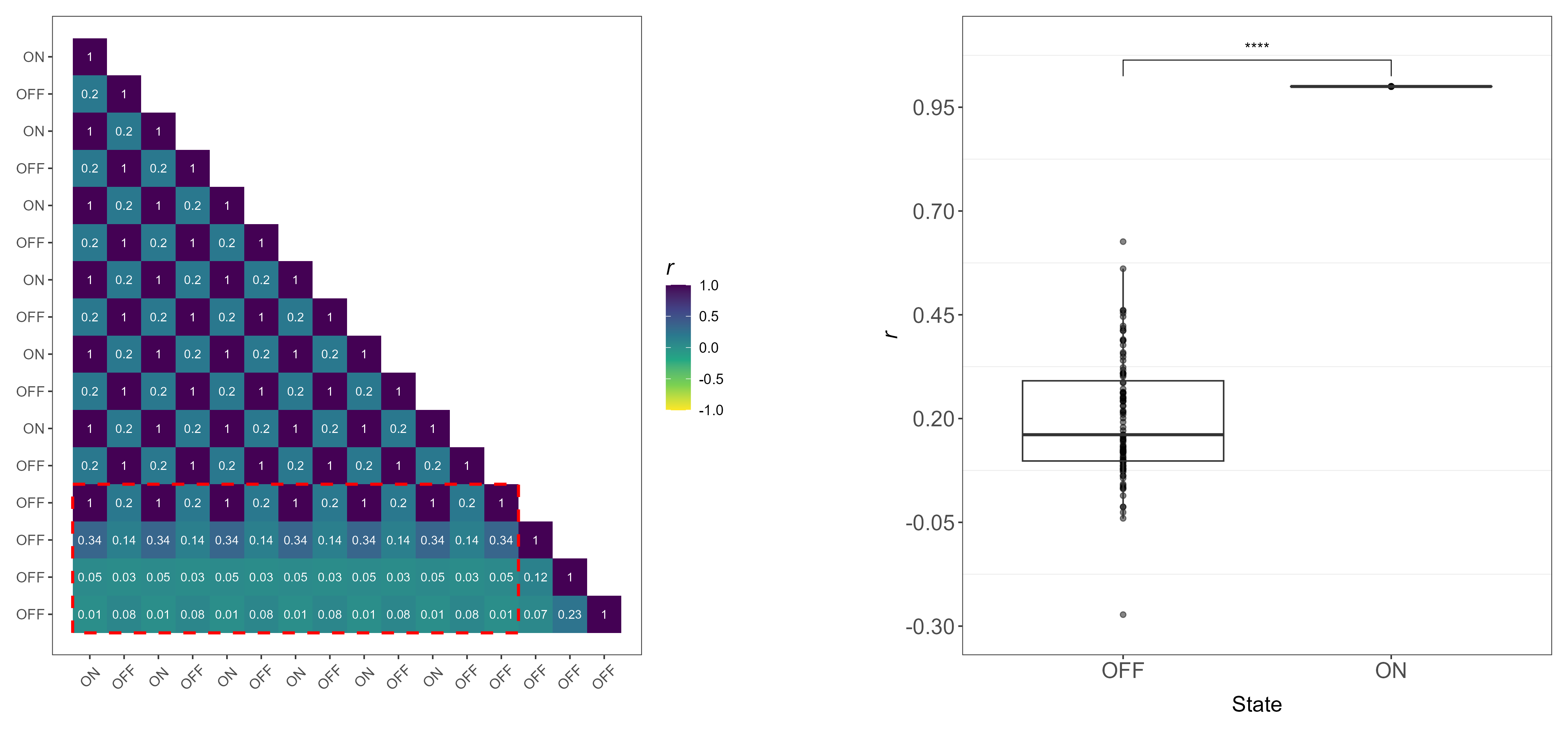}
    \caption{Fading memory of the reservoir. \textbf{Correlation matrix} (left): Correlations between the state of the reservoir on the last 12 time steps of training and four time steps into the future without further stimulation (the fading memory), averaged over 100 simulations. Time progresses from top to bottom along the y-axis and from left to right along the x-axis. The red, dashed box shows the correlation coefficients between each fading memory time step and the past 12 time steps of training. Higher correlations between fading memory states and training states indicate what input the network anticipated. The first row in the red box shows that the reservoir returned to a state most similar to past occurrences of "ON," indicating that the reservoir anticipated the next periodic phase even in the absence of the oscillatory stimulus. After this, the reservoir settled into a state unlike either oscillatory state as activation died out. \textbf{Next state anticipation} (right): The correlation coefficients between the reservoir state on the first time step of fading memory (corresponding to when "ON" would have appeared) and reservoir states from the last 12 time steps of training over 100 simulations. On average, the reservoir was more highly correlated with previous states associated with "ON" compared to "OFF," even in the absence of further input, indicating that the reservoir correctly anticipated the oscillatory signal one time step into the future.}
    \label{fig:correlations}
\end{figure*}

We assessed the fading memory of the network by correlating reservoir states after input cessation with states during the training phase. Strong correlations between post-input and prior training states would reflect that the reservoir's internal dynamics preserved traces of the periodic input, enabling the system to carry forward temporally-structured information through internal regulation alone.

We correlated the activation vector of the reservoir for four time steps after input was removed with the last 12 time steps of training. Figure \ref{fig:correlations} shows the correlation matrix averaged over 100 simulations of the model. Time is indicated along the y-axis from top to bottom and along the x-axis from left to right. The comparison of interest between the fading memory states and prior states when the signal was present is highlighted by the red, dashed box.

The reservoir carries the signal forward for an additional time step (when an "ON" stimulus would be expected) after which the reservoir states do not distinctly correlate with either the "ON" or "OFF" state as activation in the network fades to zero. This is indicated by the fact that the state of the reservoir activity on the first time step of the fading memory more highly correlated with past reservoir states when an "ON" signal was presented (\textit{r}=1) compared to past reservoir states when an "OFF" signal was presented (\textit{r}=0.18).

Figure \ref{fig:correlations} also shows the correlation values between the first time step of the fading memory and the two possible states (ON/OFF) of the reservoir from the last 12 time steps of the training regime over 100 simulations. A one-sided t-test using Fisher Z transformed correlations indicated that the reservoir state was more highly correlated with past states when "ON" was presented compared to past states when "OFF" was presented, \textit{t}(629.67) = 173.49, \textit{p} $<$ 2.2e-16. Lastly, there appears to be greater variability in correlation values with the incorrect "OFF" prediction ($\mu$ = 0.198, $\sigma$ = 0.144) compared to the correct "ON" prediction ($\mu$ = 0.999, $\sigma <$ 0.001). 

The next time step anticipation exhibited by the basal reservoir resembles the behaviour of a trained slime mould: it retracts in response to exposure to light or heat, resumes its prior state upon removal of the stimulus, and subsequently retracts again in  anticipation of the stimulus's return -- even in the absence of its actual reintroduction \citep{boussard2021adaptive}.

\section{Discussion}

We presented a basal reservoir -- a non-neural, adaptive system inspired by \textit{Physarum polycephalum} -- that demonstrates anticipatory behaviour through local, unsupervised, homeodynamic regulation without the need for explicit spiking mechanisms or readout layers. Our results show that the reservoir, driven solely by local metabolic regulation and energy balancing, encodes temporal patterns from environmental inputs, allowing for simple predictive behaviour via internal dynamics alone. Following \cite{falandays2021prediction}, we analysed the fading memory of the reservoir by examining how reservoir states after input removal correlated with earlier training states. Strong correlations indicated that internal dynamics preserved the temporal structure of past inputs even without ongoing external stimulation, demonstrating a form of pattern completion through time. 

Although the anticipatory effect fades quickly -- often lasting only a single time step -- this may reflect the model’s minimal design, which captures core features of non-neural adaptive dynamics without added complexity. Interestingly, similarly brief anticipation is observed in \textit{Physarum polycephalum}. In the study by \cite{saigusa2008amoebae}, the organism, on average, anticipated just over half a cycle following periodic stimulation, with most responses persisting for only one or two cycles and many trials showing no anticipation at all. This limited persistence suggests that brief internal carryover may be a biologically grounded feature of non-neural anticipatory systems, rather than a shortcoming of the model.

Our results also echo a recent study by \cite{pla2025minimal} that shows that anticipation can emerge in systems without nervous systems or learning algorithms, such as synthetic genetic circuits. While Pla-Mauri and Solé demonstrate that anticipation can arise in a minimal genetic circuit that detects trends by comparing fast and slow responses to an input signal, and \citet{pershin2009memristive} show a similar effect using a memristive LC circuit, both approaches rely on fixed architectures. By contrast, our model shows that anticipation can also emerge through recurrent homeodynamics in spatially distributed systems. Their models respond to trends \citep{pla2025minimal} or recall memory through engineered circuit dynamics \citep{pershin2009memristive}; ours learns and internally replays temporal structure. Together, these approaches highlight that anticipatory behaviour in non-neural systems can arise from diverse mechanisms, including fixed circuit designs and adaptive, self-organizing dynamics.

While homeostasis supports internal stability, our model goes further and implements what \cite{bich2016role} call \textit{adaptive regulation}: each node not only balances its energy level but also reconfigures its incoming connections and target values in response to local conditions. \cite{bich2016role} distinguish between \textit{dynamic stability} and \textit{adaptive regulation}, arguing that minimal cognition arises only in systems capable of the latter -- those that generate endogenous accommodations to their environment rather than merely resisting change. In this sense, our model qualifies: it adaptively reshapes its internal structure through distributed interactions with its environment. Our model also aligns with \citet{keijzer2013nervous}, who suggest that early nervous systems evolved not for input–output processing but for coordinating internal activity across tissues. Similarly, our model performs global pattern completion through local rules, without relying on a readout layer.

\cite{seoane2019evolutionary} hypothesizes that reservoir computing architectures may emerge in nature when learning is cheap and tasks are variable, pointing to examples such as cortical microcolumns, spinal central pattern generators, and gene regulatory networks as potential cases. However, he also cautions that reservoir computers may be evolutionarily unstable: as systems scale up or specialize, general-purpose reservoirs may be pruned or reorganized into more efficient, task-specific networks. Our model explores an alternative possibility, where computation is not externalized into a readout layer but remains enacted through the system’s internal dynamics. Rather than being inferred or decoded, anticipatory behaviour in our basal reservoir arises directly from local allostatic regulation. This offers an alternative to the conventional reservoir computing framework, suggesting that in metabolically constrained organisms reservoir-like computation may be transient and embodied, emerging only to the extent that it supports ongoing adaptive regulation. This also echos \cite{turner2019homeostasis}, who argues that homeostasis itself can be a cognitive process, by enabling living systems to reduce stress and maintain internal order in the face of uncertainty. Our model consolidates this view: nodes regulate local imbalances without computing prediction errors or receiving rewards, yet, collectively exhibit memory and anticipation. 

Another evolutionary perspective is offered by Levin, who outlines a trajectory from basic homeostasis to more complex cognitive capacities -- beginning with stress-reduction, followed by information-seeking behaviours (infotaxis), and eventually giving rise to memory, prediction, and modular goal pursuit in multicellular organisms \citep{levin2019computational}. While our model does not engage in active information seeking, it demonstrates that anticipatory behaviour and temporal memory can still emerge through internal regulatory dynamics alone.

Although the task we demonstrated here is deliberately simple, the model offers a first step toward understanding how adaptive regulation can support temporal pattern completion in non-neural systems. Although inspired by \textit{Physarum polycephalum}, our model is not intended as a detailed physiological account. Rather, it serves as a minimal abstraction to explore how anticipatory behaviour might emerge from local homeodynamic principles. This conceptual framing opens several promising directions for future work. One is to explore the limitations of this state-continuous, non-spiking architecture in comparison to the spiking model of \cite{falandays2021prediction, falandays2024resonance}. To address this, we plan to evaluate the model’s ability to handle more complex temporal patterns, such as nested periodicities, assess how far into the future anticipation can persist, and examine how structural features — such as network geometry or local connectivity — influence memory and adaptive responses.

Another direction is to incorporate more biologically grounded mechanisms inspired by \textit{Physarum polycephalum}, such as calcium-based dampening and oscillatory feedback dynamics, to better reflect the physiological basis of its adaptive behaviour. These features may allow a basal reservoir to regulate activation levels in a more physiologically realistic manner -- potentially offering a continuous analogue to spiking and capturing more of the organism’s dynamics. We hope this model offer a minimal starting point for exploring how learning and anticipation can emerge in simple, non-neural networks through internal regulation and distributed adaptation.

\section{Acknowledgements}
We thank Dr. Michael J. Spivey for his valuable feedback during the preparation of this manuscript. Part of this research was performed while the authors were visiting the Institute for Pure and Applied Mathematics (IPAM), which is supported by the National Science Foundation (Grant No. DMS-1925919). This work was also supported by an Institute for Humane Studies Fellowship (Grant No. IHS018559), supporting Polyphony Bruna, and by funding from the Knut and Alice Wallenberg Foundation (Grant No. 2023.0420), the Swedish-American Fulbright Commission, and the G.S. Magnuson Foundation (Grant No. 2024-0067), supporting Linnéa Gyllingberg.
\footnotesize
\bibliographystyle{apalike}
\bibliography{sample} 

\end{document}